# Superconducting dome by tuning through a Van Hove singularity in a two-dimensional metal

Wen Wan[1,], Rishav Harsh[1], Paul Dreher[1,], Fernando de Juan[1,2] and Miguel M. Ugeda[*,1,2,3]

[1]*Donostia International Physics Center (DIPC), Paseo Manuel de Lardizábal 4, 20018 San Sebastián, Spain.*

[2]*Ikerbasque, Basque Foundation for Science, 48013 Bilbao, Spain.*

[3]*Centro de Física de Materiales (CSIC-UPV-EHU), Paseo Manuel de Lardizábal 5, 20018 San Sebastián, Spain.*

*\* Corresponding author: mmugeda@dipc.org*

## *Abstract*

*Chemical substitution is a promising route for the exploration of a rich variety of doping- and/or disorder-dependent collective phenomena in low-dimensional quantum materials. Here we show that transition metal dichalcogenide alloys are ideal platforms to this purpose. In particular, we demonstrate the emergence of superconductivity in the otherwise metallic single-layer $TaSe_2$ by minute electron doping provided by substitutional W atoms. We investigate the temperature- and magnetic field-dependence of the superconducting state of $Ta_{1-\delta}W_\delta Se_2$ with electron doping ($\delta$) using variable temperature (0.34–4.2 K) scanning tunneling spectroscopy (STS). We unveil the emergence of a superconducting dome spanning $0.003 < \delta < 0.03$ with a maximized critical temperature of 0.9 K, a significant increase from that of bulk $TaSe_2$ ($T_C = 0.14$ K). Superconductivity emerges from an increase of the density of states (DOS) as the Fermi surface approaches a van Hove singularity due to doping. Once the singularity is reached, however, the DOS decreases with $\delta$, which gradually weakens the superconducting state, thus shaping the superconducting dome. Lastly, our doping-dependent measurements suggest the development of a Coulomb glass phase triggered by disorder due to W dopants.*

*Introduction*

The isolation and manipulation of atomically-thin crystals have recently enabled the investigation of wealth of exotic electronic phenomena. A remarkable example is the case of superconductivity (SC) in transition metal dichalcogenide (TMD) monolayers, where the strong spin-orbit coupling (SOC), together with the lack of inversion symmetry, triggers the emergence of unconventional superconducting properties such as the Ising pairing and symmetry-allowed triplet configurations[1–5]. Furthermore, most of the bulk TMD metal counterparts are superconducting, which enables interrogation of the effects of dimensionality and interlayer coupling. For example, the TMD metals 2H-$MX_2$ (M = Nb, Ta and X = S, Se) are intrinsic superconductors in bulk, but exhibit disparate behavior as they are thinned down to the monolayer. In $NbX_2$ monolayers, superconductivity shows significant weakening ($NbSe_2$)[6,7] and even disappearance ($NbS_2$)[8] with respect to bulk. In contrast, electron transport experiments in ultrathin films of $TaX_2$ (even reaching the monolayer in $TaS_2$) have revealed significant increase of the critical temperature ($T_C$)[9–13]. In $TaSe_2$, ionic-gating measurements revealed a substantial increase of $T_C$ = 1.4 K of nm-thick films (~5 layers) as compared to the $T_C$ = 0.14 K of bulk[10]. However, ultrathin films of TMD metals are highly susceptible to degradation in ambient conditions, which affects the intrinsic properties of collective electronic states (SC and CDW). Unfortunately, experimental work exploring the thickness dependence of SC in inert controlled environment is still scarce[6,12]. This leads to disparate qualities among the TMDs, which often make the results hardly comparable. Therefore, a coherent picture about the impact of dimensionality on the SC state has not yet been achieved for this family of layered materials.

In parallel to the exploration of the intrinsic ground state in correlated 2D systems, there has recently been an increasing interest in tuning their many-body electronic states. A successful approach here is the use of ionic gating, which has demonstrated its effectiveness to reversibly manipulate the collective phases of 2D TMDs[14–16]. An alternative method to tune the electronic properties of a layered material is by chemical doping. However, this approach has been rarely used to manipulate many-body states[17,18], and most effort so far has focused on tuning the bandgap and mobility of TMD semiconductors[19–23]. Furthermore, chemical-doping strategies enable to interrogate the robustness and fundamental properties of many-body states in the presence of disorder, which remain largely unexplored due to the lack of suitable platforms.

In this work, we demonstrate that a monolayer of TaSe$_2$ does not hold superconductivity down to 340 mK (using samples unexposed to air) by using variable-temperature (0.34 – 4.2 K) scanning tunneling spectroscopy (STS). Furthermore, we induce superconductivity in this 2D material by electron doping using W atoms given the proximity of the Fermi level (E$_F$) to an empty van Hove singularity (vHs) in its DOS. First, our spatially resolved STS measurements confirm that W atoms are embedded in the TaSe$_2$ lattice acting as electron donors. We subsequently probe the low-energy electronic structure of lightly doped TaSe$_2$ monolayers and unveil the emergence of a superconducting dome on the temperature-doping phase diagram. Optimized superconductivity develops for a W concentration of 1.8% with T$_C$ ~0.9 K, a significant increase from that of bulk (T$_C$ = 0.14 K). The SC dome reflects the variations in the available electrons for pairing caused by the crossing of a vHs in the DOS spectrum, as the layer is electron doped. Lastly, we identify the formation of a Coulomb glass phase related to disorder and boosted by W dopants.

## *Results and discussion*

### *Pristine monolayer TaSe$_2$*

Our STM/STS experiments were carried out on Ta$_{1-\delta}$W$_\delta$Se$_2$ alloys in the dilute regime (0 < δ < 0.07) grown on bilayer graphene (BLG) on 6H-SiC(0001) via molecular beam epitaxy (see methods). We use graphene as a substrate since it plays a negligible role in the electronic structure of TMD monolayers, including the superconductivity[24]. Single-layer TaSe$_2$ retains the characteristic 3×3 CDW order shown in bulk (see inset fig. 1a), which induces a gap-like feature in the DOS of 2Δ$_{CDW}$ ≈ 12 meV around E$_F$ (see supplementary information), in agreement with previous experiments[25]. First, we probe the existence of superconductivity in monolayer TaSe$_2$ (δ = 0). Figure 1b shows the out-of-plane magnetic field (B$_\perp$) dependence of the low-energy electronic structure (± 3 meV) of the monolayer at T = 0.34 K. For B = 0 T, the differential conductance (dI/dV ∝ DOS) shows a pronounced dip in the DOS of width ω ~0.6 meV at E$_F$ which is, in principle, compatible with superconductivity. However, this dip is an electronic feature that remains unperturbed as B$_\perp$ is increased, as seen in the dI/dV spectra subsequently taken at $B = 3$ T and $B = 5$ T. To quantify this observation, we show the evolution of the width ($\omega$) and depth of the dip ($d$) with $B_\perp$ in fig. 1c. Both magnitudes remain roughly constant within our resolution (see methods). In summary, the insensitivity of the dip with $B_\perp$ allows us to rule out the existence of superconductivity in monolayer $TaSe_2$ down to 0.34 K. The origin of the dip will be discussed later with additional disorder-dependent (W concentration) STS data.

The absence of superconductivity found in monolayer $TaSe_2$ is in contrast to transport experiments carried out in thin-films (multilayers) of $TaSe_2$, which report $T_C \sim 1$ K[9,10]. A plausible origin of these disparate results is the sensitivity of the SC state to oxidation as the TMD approaches the 2D limit as all the previous experiments used samples exposed to ambient conditions. Another possibility is that the superconductivity follows a non-monotonic evolution from bulk down to the monolayer limit. Although unlikely, we cannot exclude this possibility and, therefore, detailed studies are required to fully understand the evolution of the superconductivity in $TaSe_2$. We emphasize, however, that the role of the BLG substrate in the superconductivity is negligible, as we previously demonstrated by showing identical properties of single-layer $NbSe_2$ on BLG and h-BN[24]. The evolution of the superconductivity in $TaSe_2$ with crystal thickness departs from $TaS_2$, which undergoes a significant increase of $T_C$ in the monolayer (3.5 K) with respect to the bulk (0.8 K).

*Electron-doped TaSe₂*

The absence of superconductivity in monolayer $TaSe_2$ raises the question whether electron doping could trigger its emergence. Electron-doping in $TaSe_2$ is expected to increase the available electrons at $E_F$ to pair due to the proximity to a vHs in the DOS[26–28]. To this purpose, individual W atoms were embedded as substitutional atoms in the Ta plane during the layer growth. W has one more valence electron than Nb and, therefore, acts as electron donor (see supplementary information for the electronic characterization of the W dopants). The resulting monolayer is an aliovalent $Ta_{1-\delta}W_\delta Se_2$ alloy with a dilute W concentration ($\delta$) that can be precisely characterized via STM imaging. As shown in fig. 2, for W doping levels as low as 2.5% ($\delta = 0.025$) superconductivity develops. Figure 2a shows the evolution of the DOS in $Ta_{0.975}W_{0.025}Se_2$ with $B_\perp$ (see methods for details on the $B_\perp$-dependent STS measurements). At zero field, the DOS also shows a similar dip centered at $E_F$. However, in contrast to the pristine TaSe$_2$ case (fig.1), this dip is now highly susceptible to $B_\perp$, and its depth gradually decreases as $B_\perp$ increases up to an upper critical field $B_{C_2} = 0.7$ T, a value beyond which the DOS remains unchanged (see supplementary information for the definition of critical values $B_{C_2}$ and T$_C$). Furthermore, the dip now shows peaks at its edges at B = 0 T that gradually disappear with $B_\perp$, which are consistent with superconducting coherence peaks. Figure 2b shows the set of dI/dV($B_\perp$) spectra of fig. 2a normalized to dI/dV(0.75 T). As seen, a clear DOS dip centered at E$_F$ evolves as $B_\perp$ is decreased. Since this sensitivity against $B_\perp$ is compatible with superconductivity, we fit these dI/dV spectra to a BCS gap $\Delta_{BCS}$ (see supplementary information for fitting details), and plot them as a function of $B_\perp$ (fig. 2c, red dots). This figure shows the same measurements in another two regions of the same sample (pink and orange dots).

To confirm that the dip feature emerging in the normalized dI/dV spectra corresponds to a superconducting gap, we also study its temperature dependence in the same sample. Figure 2d show a series of dI/dV spectra acquired consecutively at different temperatures. Similarly, the dip with bound peaks decreases as the T increases up to 0.9 K, where it remains nearly unchanged. This evolution is better observed in the normalized spectra to dI/dV(0.9 K) in Fig.2e. We fit the normalized dI/dV(T) spectra to a SC gap following the same procedure, and fig. 2f shows the evolution of $\Delta_{BCS}$ with T. As seen, $\Delta(T)$ follows a BCS-like temperature dependence (black curve) with $\Delta(0) \approx 0.2$ meV.

Once the emergence of superconductivity in W-doped TaSe$_2$ monolayers is established, we further investigate its evolution with W concentration, which is linearly proportional to the electron doping. We have studied 22 samples with different δ to cover the the range $0 < δ < 0.07$. For each sample, we measure both the T and $B$ dependence of the low-lying electronic structure, as previously described (Fig.2). Figure 3 summarizes the evolution of the SC gap, the upper critical field ($B_{C_2}$) and T$_C$ as a function of δ. The plotted values represent averaged values among all the regions studied for each δ (see supplementary information). As seen, a superconducting dome spanning $0.003 < δ < 0.03$ is found with a maximal T$_C \approx 0.9$ K for δ = 0.018. Superconductivity develops for W concentrations as dilute as 0.5% (δ = 0.005). This phase diagram reveals two additional interesting features. First, the superconducting state in the monolayer is significantly more robust against $B_\perp$ with upper critical fields as large as $B_{C_2} \approx 1.1$ T with respect to the bulk, where $B_{C_2}$ is in the mT range. Second, at the optimum W concentration for superconductivity, the SC gap reaches a value of $\Delta_{BCS} \approx 0.18$ meV, which leads to a lower bound for the ratio $2\Delta(0)/k_B T_C \geq 4.6$, a value significantly larger than the predicted by the BCS theory (3.53). This indicates that the pair-coupling interaction is strong in the W-doped TaSe$_2$ monolayer. This enhanced value is nearly coincident with that of single-layer NbSe$_2$ ($\Delta(0) = 0.4$ meV, $T_C \approx 2$ K). Lastly, the Ginzburg-Landau coherence length estimated for this 2D alloy at optimum doping is $\xi_{GL}$(T = 0 K) =13.8 nm, also very close to that for single-layer NbSe$_2$ ($\approx 18$ nm)[29,30].

Next, we discuss the origin of the superconductivity in electron-doped TaSe$_2$. The band structure of monolayer TaSe$_2$ has a saddle point near the $E_F$, around mid-way in the Γ-K direction, which leads to a logarithmic divergence in the DOS known as a Van Hove singularity. SOC splits this vHs by about 0.23 eV according to DFT calculations[26–28], resulting in one peak slightly above $E_F$ and one below it. This is also supported by ARPES experiments, which observe only one band along Γ-K[25]. As we dope electrons in Ta$_{1-\delta}$W$_\delta$Se$_2$ by increasing δ, $E_F$ is tuned through the vHs, where superconductivity should be enhanced due to the increased DOS. This is therefore a natural explanation for the observed emergence of the superconducting dome, similar to those found in 3D materials where the vHs is tuned by chemical doping[31] or strain[32].

To estimate the amount of W needed to reach the vHs, an accurate calculation of the DOS, the vHs and $E_F$ positions is required. Since this value is not accurately reported in the literature, we use a tight-binding model with hoppings up to 5 nearest neighbors and SOC, chosen to reproduce the main features of the DFT bands of Refs.[26–28]. The resulting band structure and corresponding DOS are shown in fig. 4. The distance from $E_F$ to the upper vHs peak is taken to be 5% of the vHs SOC splitting, which is within the values inferred from Refs.[26–28]. This model is then used to compute the amount of electron doping needed to reach the vHs peak, which we obtain to be 0.015 electron/cell. This corresponds to a value of δ = 0.02, in agreement with the dome maximum location.

*Coulomb glass phase*

A remaining question is the origin of the dip at $E_F$ intrinsically present in pristine and W-doped TaSe$_2$. To gain knowledge here, we have analyzed the DOS within ± 1 mV for non-superconducting samples. Figure 5a shows three typical dI/dV spectra for different W concentrations. The dip within ± 1 mV shows a gradual increase of its width with δ accompanied by a change in shape. In particular, ω shows a non-linear increase with W concentration (plot in fig.5b, where ω represents the averaged ω over several different regions and/or samples). Furthermore, the shape of the dip evolves qualitatively towards a linear shape with W concentration (fig. 5c).

At δ = 0, a plausible origin for this intrinsic dip could be a CDW-induced partial gap within the main CDW gap (see supplementary information). However, the doping dependence of the dip would not conform to a CDW origin because chemical doping introduces disorder scattering, which would tend to suppress the CDW. Since disorder is present even in the undoped samples, we rather surmise that the dip is a disorder effect: the combination of small disorder and Coulomb interactions leads to a logarithmic dip of the DOS near $E_F$[33,34] (see fit in fig. 1d). When disorder is further increased by doping, in 2D this dip evolves into power-law known as Coulomb gap, which widens with disorder. While the power is often quoted to be linear, numerical studies often deviate from this value[35,36]. Nevertheless, these observations are consistent with the emergence of a Coulomb gap driven by disorder. Since disorder is not expected to affect superconductivity due to Anderson's theorem, our global picture of $Ta_{1-\delta}W_\delta Se_2$ is that W substitution generates superconductivity as the nominal DOS is tuned through the vHs, while also increasing disorder so that a Coulomb gap develops at larger disorder strengths.

*Discussion*

In summary, we prove the existence of superconductivity in a 2D metal with strong SOC where the Fermi surface is tuned at a saddle point in the band structure. Superconductors with such a band structure are candidates to host chiral/helical superconductivity and magnetic orders[37–41], and few of them have been identified in nature[42,43]. Our results open the doors to further experimental investigation in this type of alloys in search for signatures of topological superconductivity.

Lastly, the absence of superconductivity in a nearly isolated high-quality monolayer of $TaSe_2$ changes the picture of dimensionality effects on superconductivity in TMDs and calls for further theoretical efforts. Furthermore, our work highlights TMD alloys as ideal playgrounds for the study of a variety of disorder-driven electronic phase transitions in purely 2D systems, which have remained largely unexplored due to the lack of suitable platforms.

## Methods

### Sample growth

Single-layer $Ta_{1-\delta}W_{\delta}Se_2$ samples ($0 < \delta < 0.07$) were grown on BLG/SiC(0001) substrates. The details for preparing BLG on SiC(0001) can be found in ref. 24. For the growth of the TMD layer we co-evaporated high-purity Ta (99.95%), W (99.99%) and Se (99.999%) in our home-made molecular beam epitaxy (MBE) system under base pressure of ~5.0 × $10^{-10}$ mbar. The flux ratio between transition metals (Ta and W) and Se is 1:30. During the growth, the temperature of BLG/SiC(0001) substrates were maintained at 570℃ and the growth rate was ~2.5 hours/monolayer. The flux of Ta was kept constant and the flux of W can be proportionally changed to control the stoichiometry δ. After the growth of the monolayer alloys, the samples were kept annealed in Se environment for 2 minutes, and then immediately cooled down to room temperature. In-situ RHEED was used for monitoring the growth process. After the growth, a Se capping layer with a thickness of ~10 nm was deposited on the surface to protect the film from contamination during transport through air to the UHV-STM chamber. The Se capping layer was subsequently removed in the UHV-STM by annealing the sample at ~300 °C for 40 minutes prior to the STM measurements.

### STM/STS measurements and energy resolution at 0.34 K

STM/STS data were acquired in a commercial UHV-STM system (Unisoku, USM-1300) operated at temperatures between 0.34 K and 4.2 K. This system is equipped with out-of-plane magnetic fields up to 11 T. For these experiments we used Pt/Ir tips previously calibrated on Cu(111) to avoid tip artifacts. The typical lock-in a.c. modulation parameters for low- and large-bias STS were ~ 30 μV (833 Hz) and ~ 5 mV (833 Hz), respectively. The dI/dV spectra shown in this work are averaged spectra from spatial grids (6×6 points) in regions of 10×10nm$^2$. The WSxM software were used for analysis and rendering our STM/STS data[44].

Since the superconducting gap structure in type-II superconductors becomes spatially dependent in the presence of a $B_\perp$-field, we probed the $B_\perp$-field dependence of the DOS near $E_F$ by carrying out nm-sized grid dI/dV spectroscopy in several locations of each sample for a given δ. The gradual weakening of the dip feature was always found in the 0.003 < δ < 0.03 range, where superconductivity develops (summarized in the supplementary figure 5).

The energy resolution of our dI/dV measurements (dI/dV ∝ LDOS) at the lowest possible operational temperature of 0.34 K has been verified by measuring Josephson supercurrents in the two-band superconductor bulk Pb(111) with a superconducting Pb tip, which reveals a broadening of Γ ~105 ± 10 μV (see supplementary information). This broadening is mostly caused by the thermal broadening at T = 0.34 K ($\Gamma_{th} = 3.5 \cdot k_B T \approx 100\ \mu eV$) and, therefore, implies that the energy resolution is mostly limited by the base temperature of the instrument.

## Data availability

The datasets generated and/or analysed during the current study are available from the corresponding author on reasonable request.

## Acknowledgements

M.M.U. acknowledges support by the the European Research Council (ERC) under the European Union's Horizon 2020 research and innovation programme (ERC Starting Grant LINKSPM no. 758558) and by the grant no. PID2020-116619GB-C21 funded by MCIN/AEI/10.13039/501100011033. R.H. acknowledges funding support for project MAGTMD from the European Union's Horizon 2020 research and innovation programme under the Marie Sklodowska-Curie grant agreement No 101033538.

## Competing interests

The authors declare no competing interests.

## Author contributions

W.W. and M.M.U conceived the experiment. W.W. carried out the MBE growth, the morphology characterization of the samples with the help of R.H. and P.D.. W.W. measured and analyzed the STM/STS data with the help of M.M.U.. M.M.U. supervised the project. F.J. provided the theoretical support. M.M.U. and F.J. wrote the paper with help from the rest of the authors. All authors contributed to the scientific discussion and manuscript revisions.

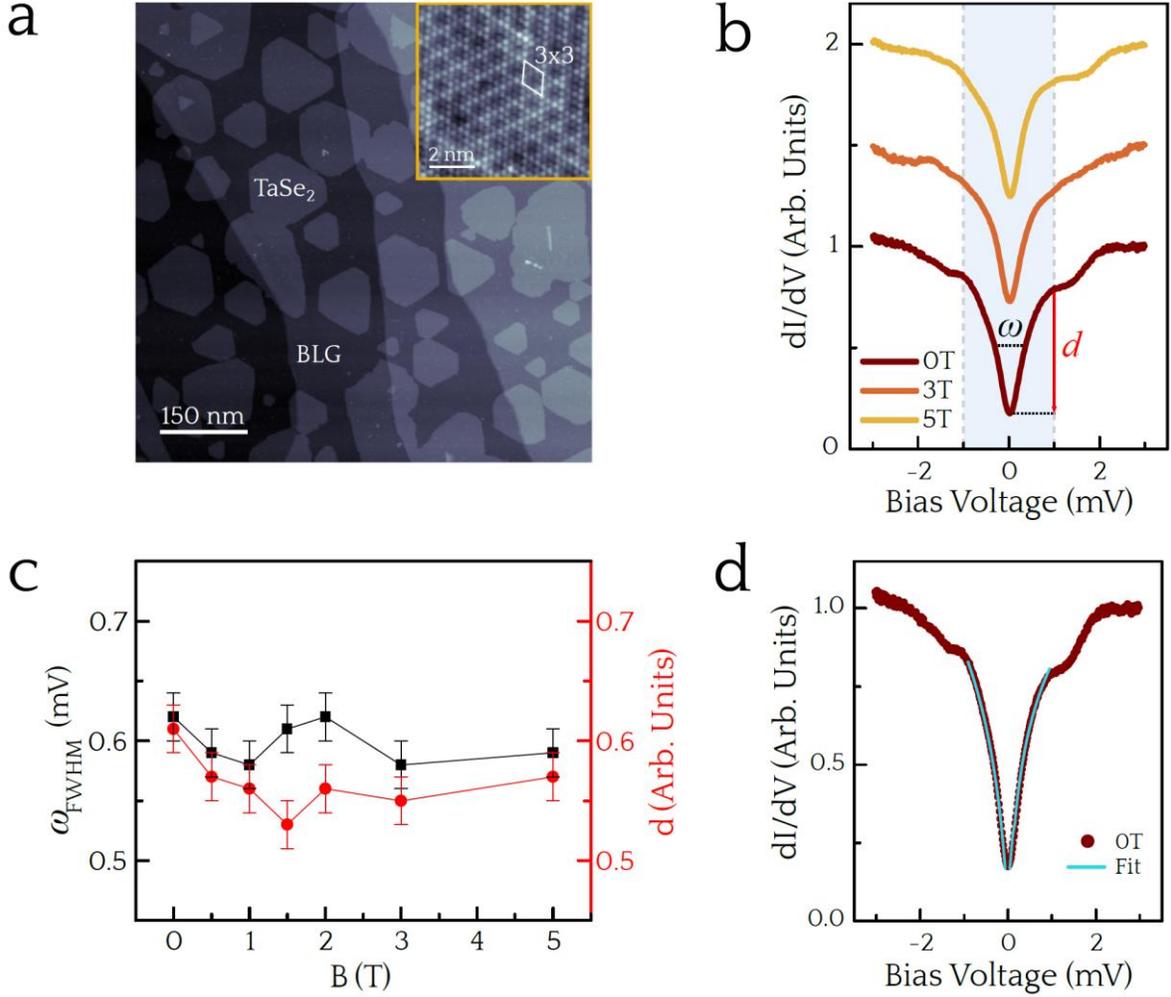

**Figure 1. Low-energy electronic structure of single-layer TaSe$_2$. a** Large-scale STM image of monolayer TaSe$_2$/BLG ($V_t$ = 2V, I = 20pA). Inset: STM image showing the 3×3 CDW ($V_t$ = 0.6V, I = 100pA). **b** $B_\perp$-dependence of the electronic structure of monolayer TaSe$_2$ (T = 0.34K). The shadowed region indicates the energy range of the inner dip. The width *(ω)* and depth *(d)* of this dip are defined with respect to the value at dI/dV(1mV). **c** The evolution of *ω* and *d* with $B_\perp$ in monolayer TaSe$_2$. The error bars represent the standard error of the mean. **d** dI/dV($B_\perp$=0T) on monolayer TaSe$_2$ (dots) and the DOS $\propto$ Ln($V$) fit within ± 1 mV (line).

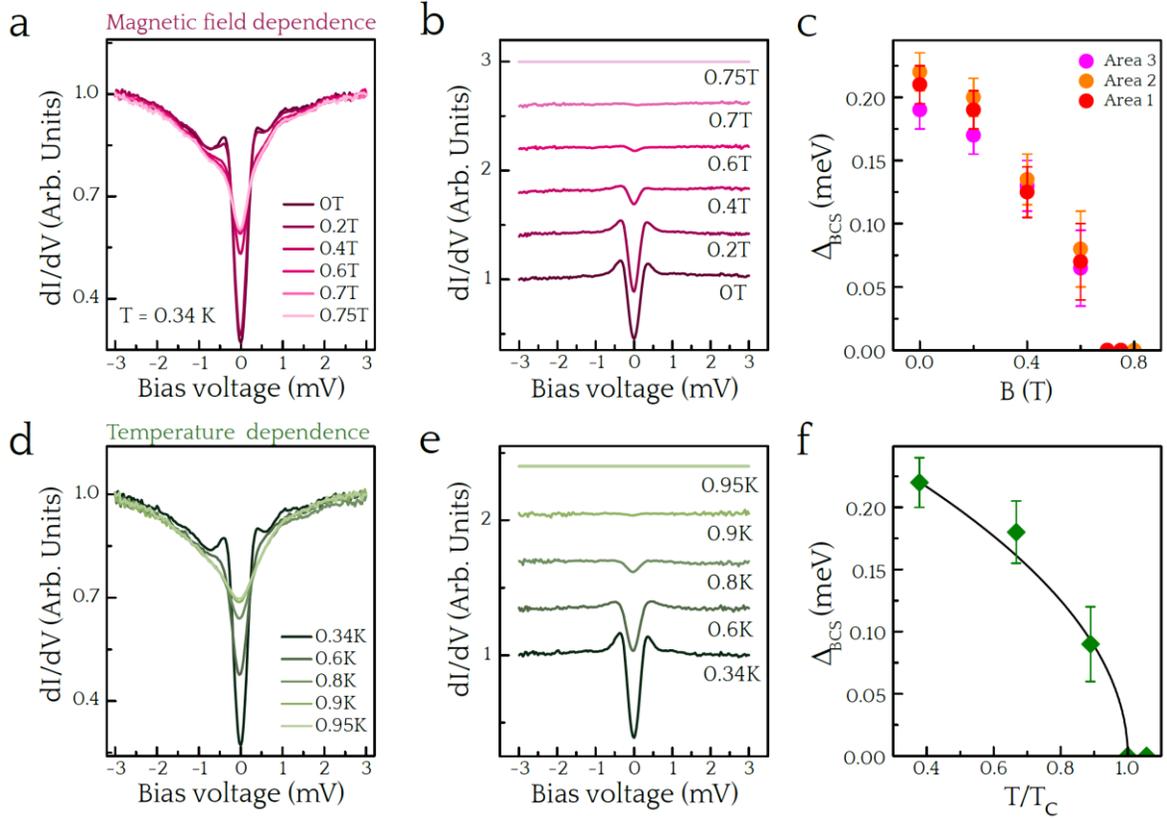

**Figure 2. Magnetic field and temperature dependence of the superconducting gap in electron-doped single-layer TaSe₂**. **a** Magnetic field-dependence of the electronic structure of $Ta_{0.975}W_{0.025}Se_2$. **b** Same dI/dV(*B*) spectra as in **a** normalized to dI/dV(0.75 T). **c** Evolution of the superconducting gap $\Delta_{BCS}$ with *B* measured in three different regions of the sample. The SC gap values have been extracted by fitting the normalized curves shown in **b**. **d** T-dependence of the electronic structure of $Ta_{0.975}W_{0.025}Se_2$. **e** Same dI/dV(*T*) spectra as in **a** normalized to dI/dV(0.95 K). **f** BCS SC fit of the dip features observed after normalization in **e**. The line shows the expected BCS temperature dependence. The error bars in **c** and **f** show the standard error of the mean.

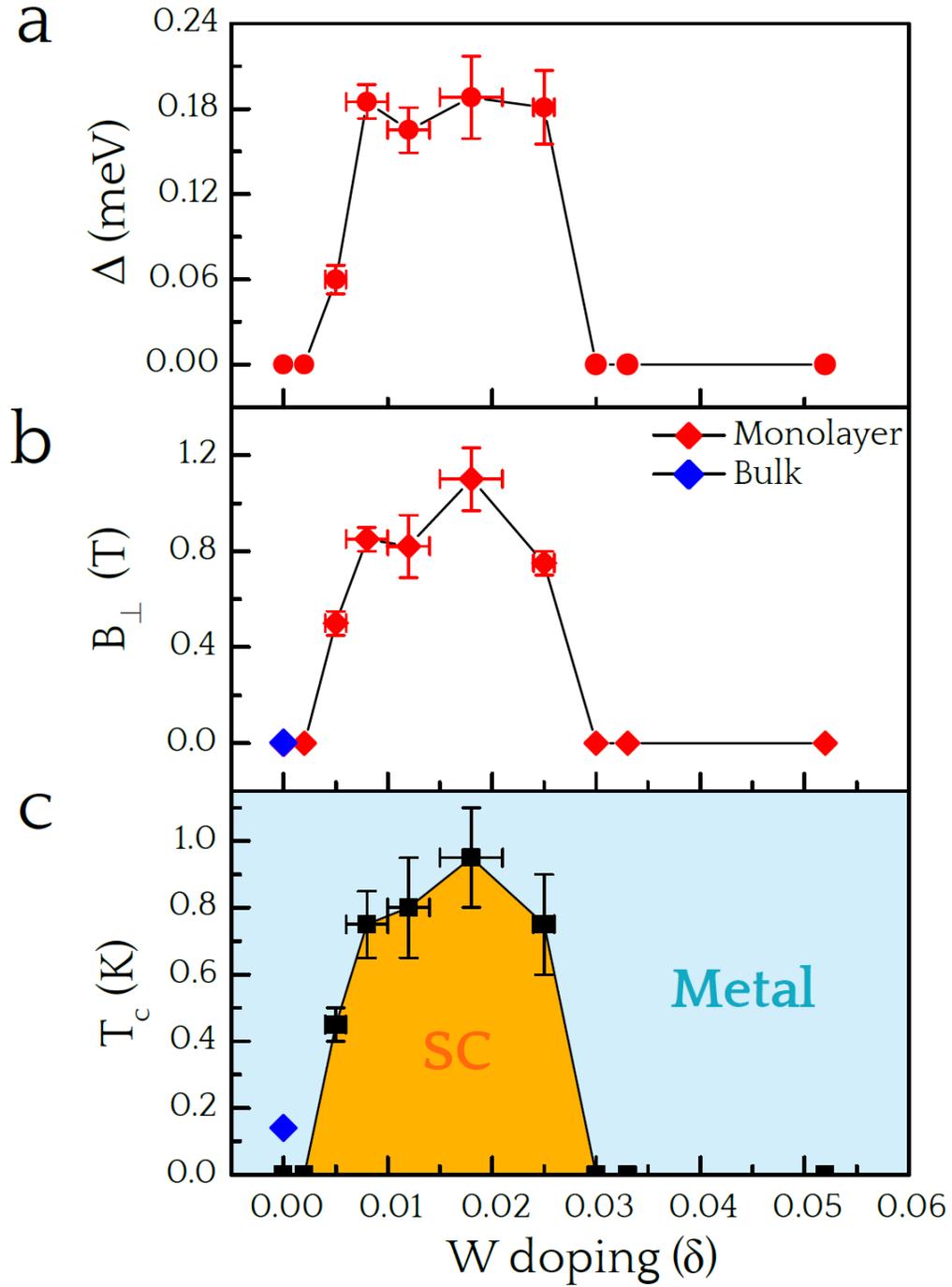

**Figure 3. Superconducting dome with electron (W) doping level.** SC gap in **a**, upper critical field in **b**, and critical temperature in **c** as a function of W doping. All the values correspond to averaged values over the regions studied for a given δ. The error bars show the standard error of the mean.

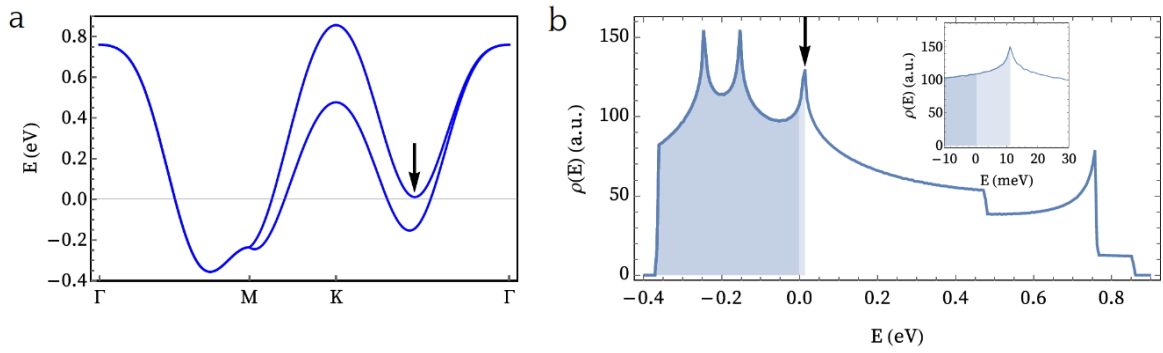

**Figure 4. Tight-binding model for TaSe$_2$. a** Band structure. The vHs near $E_F$ is marked with an arrow. **b** Calculated DOS with occupied states in dark blue. The amount of carriers needed to reach the vHs (light blue area) corresponds to δ = 0.02.

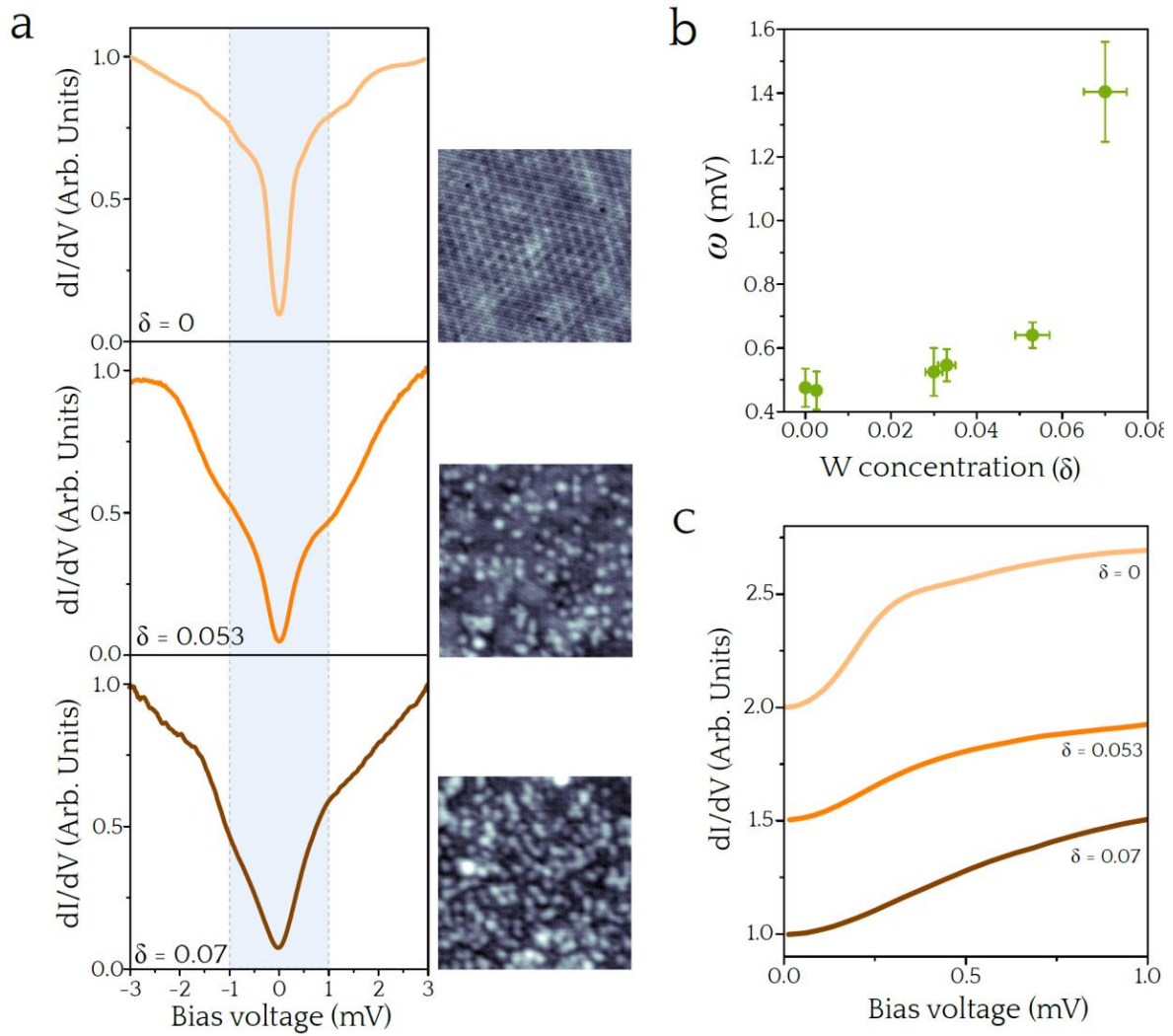

**Figure 5. Development of a disorder-induced Coulomb gap**. **a** DOS evolution of the dip in Ta$_{1-\delta}$W$_\delta$Se$_2$ monolayer with δ. The STM images on the right show the corresponding morphology of the layer. **b** Width vs. δ in the non-superconducting regime. The error bars show the standard error of the mean. **c** Zoom-in of the dI/dV spectra shown in **a** illustrating the evolution of the differential conductance towards linearity at high disorder.

# Supplementary information for

# *Superconducting dome by tuning through a Van Hove singularity in a two-dimensional metal*

Wen Wan, Rishav Harsh, Paul Dreher, Fernando de Juan and Miguel M. Ugeda[*]

*\* mmugeda@dipc.org*

## Section I: STM/STS energy resolution at 0.34 K

The energy resolution of our dI/dV measurements (dI/dV ∝ LDOS) at the lowest possible operational temperature of 0.34 K has been verified by measuring Josephson supercurrents in the two-band superconductor bulk Pb(111) with a superconducting Pb tip, as shown in supplementary figure 1. At temperatures below $T_C$, the measurements on the superconductor-insulator-superconductor (SIS) junction show relatively independent of the temperature. The source of the broadening of the Josephson tunneling peak at zero bias is solely electronic noise and is therefore ideally suited to reveal the electronic noise level of the system [1,2]. The width of the Josephson peak reveals a broadening of Γ ~100 ± 10 μV (supplementary figure 1a). The low electronic noise level of our STM system enables to spectroscopically resolve the two superconducting gaps of Pb(111), which are separated by an energy very similar to the expected thermal broadening at T = 0.34 K ($\Gamma_{th} = 3.5 \cdot k_B T \approx 100\ \mu eV$) [3] (supplementary figure 1b).

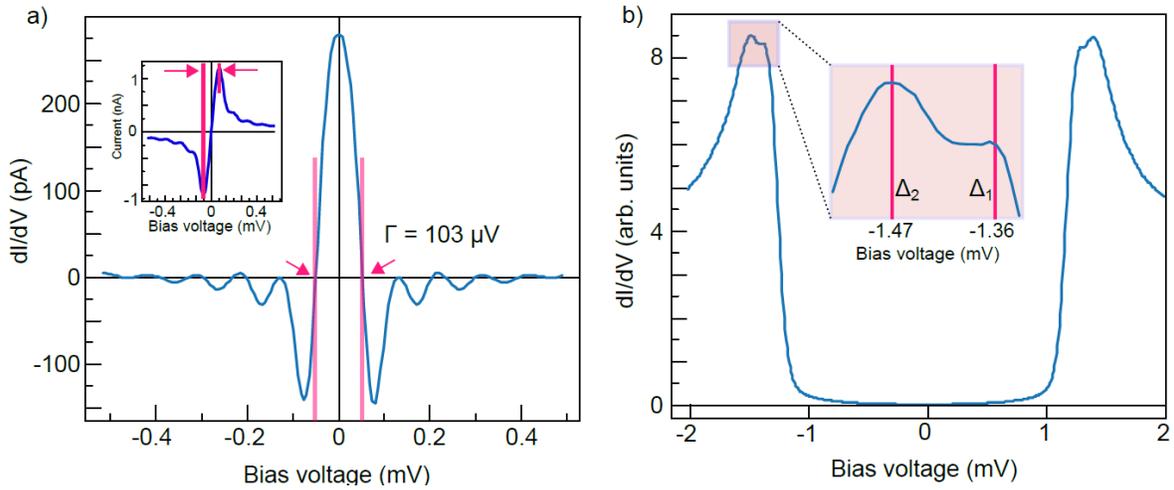

**Supplementary Figure 1. Josephson tunneling between Pb(111) and a Pb tip. a** Typical dI/dV spectrum acquired at 0.34 K through the SIS junction showing a width of the zero-bias Josephson peak of Γ ~100 ± 10 μV. The inset shows the I/V curve. **b** dI/dV spectrum taken with a normal (non-superconducting) tip on Pb(111) where the two superconducting gaps can be identified.

## Section II: Low-energy electronic structure of single-layer TaSe$_2$ (H phase)

Supplementary figure 2a shows a representative dI/dV spectrum of 1H-TaSe$_2$, which has been previously studied [4], and interpreted in terms very similar to that of 1H-NbSe$_2$ [5]. Interestingly, the low-energy electronic structure of 1H-TaSe$_2$ near E$_F$ reveals a partial gap of a width 2$\Delta_{CDW}$ = 12 ± 4 meV. This dip in the DOS has been previously related to the partial bandgap opening due to the charge density wave formation [6]. Our work identifies new internal structure within this partial gap. The most important feature is a narrow, pronounced dip in the DOS (blue region in supplementary figure 2b-c) that is the subject of the present work. The insensitivity of this feature with the magnetic field and its evolution with disorder suggests the formation of a Coulomb glass phase (see main manuscript). In addition to this feature, we also report the existence of two dim kink features indicated with green arrows in supplementary figure 2b-c of unknown origin to date.

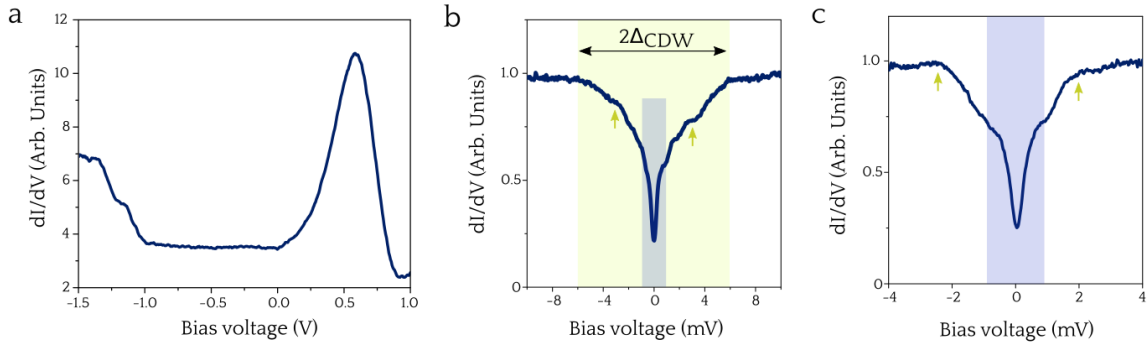

**Supplementary Figure 2**. **Electronic structure of 1H-TaSe$_2$. a** Wide-bias dI/dV spectrum of 1H-TaSe$_2$. **b** and **c** Low-energy dI/dV spectra showing the CDW gap (yellow), the robust internal partial gap (blue) and the kink features (green arrows).

## Section III: Identification of substitutional W atoms and electronic fingerprints

Supplementary figure 3a shows a typical atomically resolved STM image of single-layer Ta$_{0.994}$W$_{0.006}$Se$_2$ alloy. W substitutional atoms can be resolved individually in the atomic lattice as triangular features of equal orientation (as the one boxed). The triangular shape is due to the electronic perturbation that the W atom produces in the local DOS of the three upper neighboring Se atom [6]. Such identification is further supported by two fingerprints in the STM spectroscopy measurements. Supplementary figure 3b shows two typical dI/dV curves respectively acquired on top of a triangular feature (yellow curve) and on pristine TaSe$_2$ (blue curve). The STS taken on the triangular feature exhibits an electronic resonance centered at -0.58 eV below $E_F$ (light blue arrow), whose origin are the W-$d_z^2$ states [6]. Furthermore, this dopant also induces a rigid shift (indicated by grey arrows) of the main feature of the electronic structure for empty states, i.e., a large peak from the Ta-$d$ band whose maxima at + 0.57 ± 0.03 eV in pristine TaSe$_2$ corresponds with the energy position of this band at $\bar{\Gamma}$. This is a clear indication that W atoms induce electron filling (n-type doping) into the TaSe$_2$.

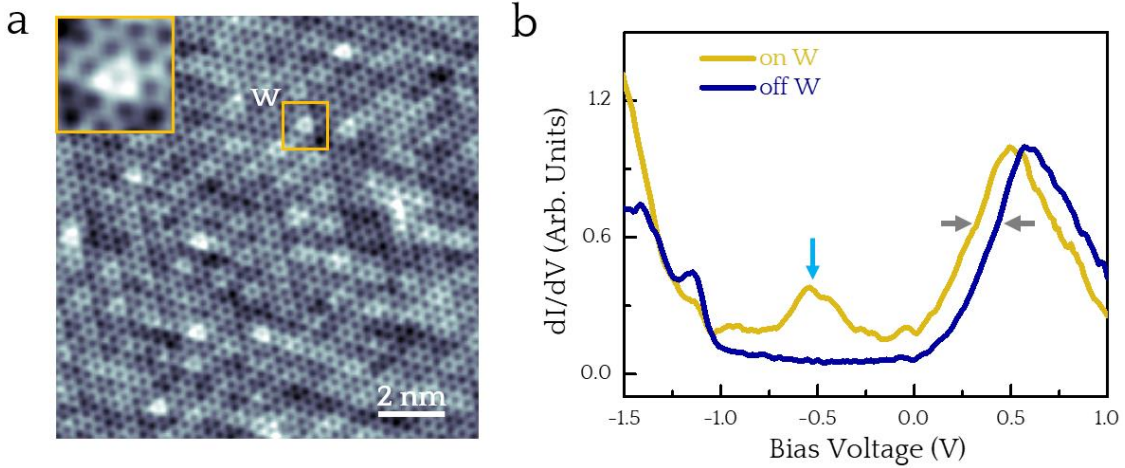

**Supplementary Figure 3. W dopants signatures. a** Atomically resolved STM image of Ta$_{0.994}$W$_{0.006}$Se$_2$ ($V_t$ = -1 V, I = 0.8 nA, T = 4.2 K). The inset show a zoom-in of the W atom boxed in the main image. **b** Wide-bias dI/dV spectra on pristine TaSe$_2$ (blue) and on one of the W atoms (yellow).

**Section IV: Determination of the W concentration (δ)**

In order to determine δ for each sample, we measured the density of W atoms in several regions (typically 30 nm x 30 nm) by W-counting from STM images. The density of W atoms was found quite homogeneous in all cases, and the quoted error of each δ (fig.3 in the main manuscript) represents the mean squared deviation from all the densities measured for a particular sample. Furthermore, we found a linear relation between the W flux from the evaporator during the growth and the estimated W density in the explored linear range.

## Section V: Determination of the critical values $B_{C_2}$ and $T_C$

To obtain the T and B-field critical values from a set of dI/dV spectra, we first normalize them using the spectrum taken at the lowest T/B that satisfies the following condition: it yields a normalized spectrum of lower T/B in which the depth of the dip at $E_F$ is smaller than to the peak-to-peak amplitude A in the ranges (± 1-3 mV), i.e., $\sigma < A$. $\sigma$ is measured from the baseline A/2 (red line in supplementary figure 4). The T/B value of this normalized dI/dV spectrum is taken as the critical value. Supplementary figure 4 shows two examples with two sets ramping the magnetic field and the temperature. For example, the dI/dV spectrum taken at 0.75 T (supplementary figure 4a) yields a normalized spectrum for B = 0.7 T that satisfies $\sigma < A$ and, therefore, B = 0.7 T is taken as the critical field for that set of dI/dV spectra.

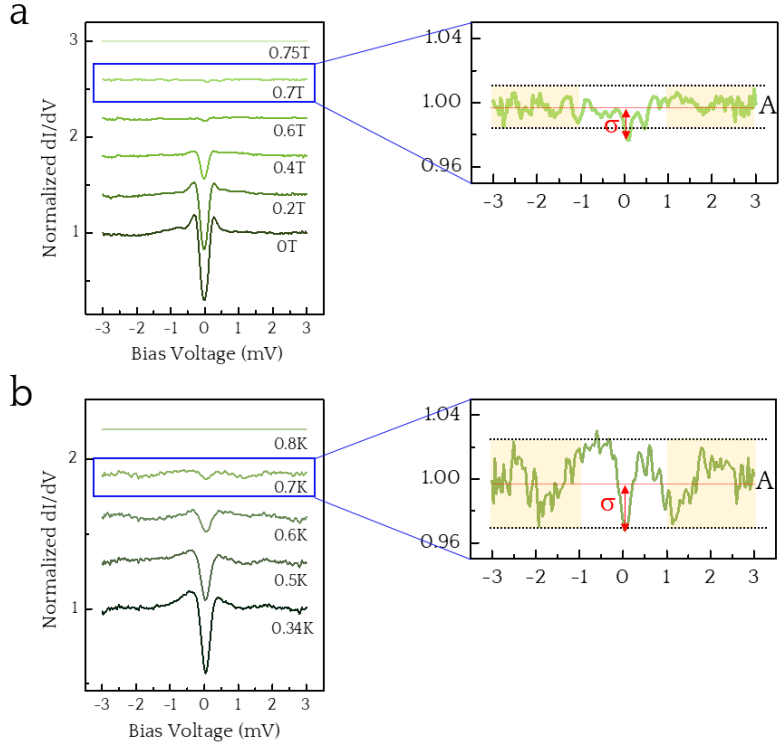

**Supplementary Figure 4**. Determination of the critical values $B_{C_2}$ and $T_C$. **a** Left, set of dI/dV spectra taken while ramping the magnetic field in doped $W_{0.008}TaSe_2$. The normalization to the spectrum at B = 0.75 T yields to a normalized spectrum taken at B = 0.7 T that satisfies $\sigma < A$ (right). Therefore, $B_{C_2}$ = 0.7 T. **b** A similar example for a set of dI/dV spectra taken while ramping the temperature.

## Section VI: The evolution of SC gap and corresponding $B_{C_2}$ on $W_\delta TaSe_2$

The evolution of SC gap and the corresponding $B_{C_2}$ with increasing of W doping ($\delta$) from 0 to 0.033 are shown in supplementary figure 6. Such evolution contains two tendencies. First, the SC generates and becomes stronger with increasing of $\delta$ up to 0.018. Second, the SC becomes weaker with keeping increasing of $\delta$ and vanishes at $\delta \geq 0.03$. The corresponding $B_{C_2}$ also follows the same evolution with SC.

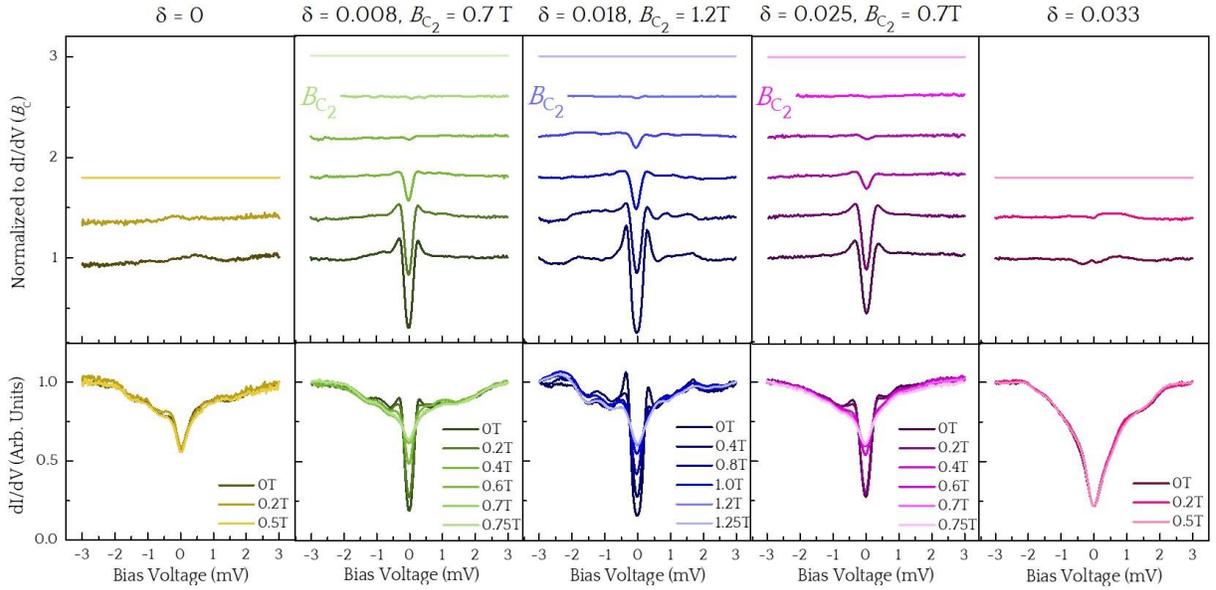

**Supplementary Figure 5. Electronic structure vs. W doping.** The evolution of SC gap and the corresponding $B_{C_2}$ with increasing of $\delta$ from 0 to 0.033. Each set of dI/dV represents the typical average electronic feature of the doped sample.

## Section VII: Fitting of the superconducting gap

The SC gap values ($\Delta(T)$) have been extracted by fitting the dI/dV spectra to the BCS expression of the quasiparticle DOS [7]:

$$\Delta_{BCS} = N_S(E, \Gamma, T) = Re\left[\frac{|E|}{\sqrt{(|E|+i\Gamma(T))^2 - \Delta(T)^2}}\right] \quad (1)$$

where $\Gamma(T)$ is a parameter associated with the lifetime of quasiparticles, although it also incorporates all broadening arising from any non-thermal sources (a.c. modulation, radio frequency noise, etc.). Thermal broadening is considered by convoluting $N_S$ with the Fermi-Dirac distribution. The mean $\Gamma$ values extracted from the fits are shown in supplementary figure 6b. As seen, $\Gamma$ does not show any particular trend with the W concentration, which likely implies that the effect of disorder on the shortening of quasiparticles' lifetime is small as compared to other non-thermal sources of broadening such as the intrinsic disorder of the layers and the instrumental electronic noise.

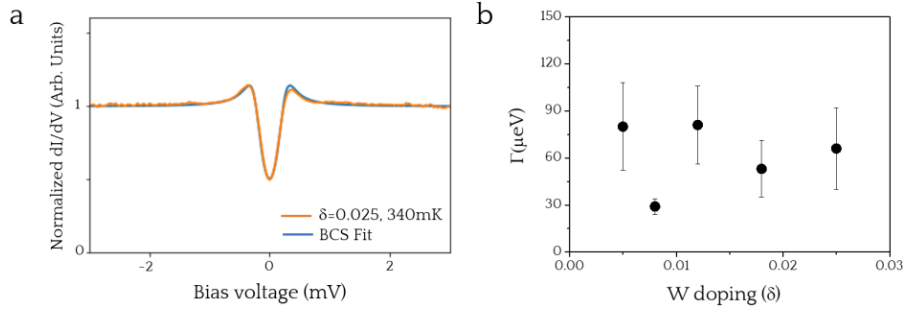

**Supplementary Figure 6**. **Quasiparticle lifetime and broadening.** **a** BCS fit of the superconducting gap measured on the sample $W_{0.025}TaSe_2$ at 340 mK. **b** Mean values of the $\Gamma$ values obtained from the fits within the superconducting regime. . The error bars represent the standard error of the mean value.

**Section VIII: DOS spectrum in single-layer TaSe$_2$**

To estimate the amount of substitutional W doping needed to reach the Van Hove singularity, we considered a 5 nearest neighbor tight binding model for a triangular lattice with SOC

$$E(\mathbf{k}) = \sum t_i \cos(\mathbf{k} \cdot \mathbf{a}_{i,n}) + t_{SOC} \sin(\mathbf{k} \cdot \mathbf{a}_{1,n}) \quad (2)$$

Where $i = 1,\ldots,5$ labels the nearest neighbors and $\mathbf{a}_{i,n}$ are the $n$ equivalent neighbor vectors for nearest neighbor $i$. The parameters were chosen to reproduce the main features of the predicted ab-initio band structure in Refs. [8–10], in particular the different band edges. The values taken were $t_i$ = (0.092, 0.227, -0.029, -0.021, -0.028), and $t_{SOC}$ = 0.073. The Fermi level was chosen such that the distance to the upper VH singularity is 5% of the splitting between upper and lower singularity, to qualitatively reproduce the placement of the Fermi level in Refs. [8–10], which slightly varies across works.